\documentclass[letter]{aa} 
\usepackage{graphicx, textcomp}
\usepackage{natbib}
\usepackage{txfonts}
 \usepackage{hyperref}
 \hypersetup{
      colorlinks   = true,
      citecolor    = violet
 }
\begin{document}

   \title{First 3\,mm-VLBI imaging of the two-sided jet in Cygnus A}
   \subtitle{Zooming into the launching region}

   \author{B. Boccardi\inst{1}
\and T.P. Krichbaum\inst{1}
\and U. Bach\inst{1}
\and M. Bremer\inst{2}
\and J.A. Zensus\inst{1}
}
\institute{Max-Planck-Institut f\"{u}r Radioastronomie, Auf dem H\"{u}gel 69, D-53121 Bonn, Germany 
\and
Institut de Radioastronomie Millim\'{e}trique, 300 rue de la Piscine, 38406 Grenoble, France  
}
\date{Received 1 March, 2016 / accepted 11 March, 2016}
 
  \abstract
   {}
{We present for the first time Very-Long-Baseline Interferometry images of the radio galaxy Cygnus A at the frequency of $86$ $\rm GHz$. Thanks to the high spatial resolution of only $\sim$200 Schwarzschild  radii ($R_{\bf S}$), such observations provide an extremely detailed view of the nuclear regions in this archetypal object and allow us to derive important constraints for theoretical models describing the launching of relativistic jets.}
{A pixel-based analysis of the jet outflow, which still appears two-sided on the scales probed, was performed. By fitting Gaussian functions to the transverse intensity profiles, we could determine the jet width in the nuclear region.}
  {The base of the jets appears wide. The minimum measured transverse width of ${\sim} (227\pm98)$ $R_{\bf S}$ is significantly larger than the radius of the Innermost Stable Circular Orbit, suggesting that the outer accretion disk is contributing to the jet launching. The existence of a faster and Doppler de-boosted inner section, powered either from the rotation of the inner regions of the accretion disk or by the spinning black hole, is suggested by the kinematic properties and by the observed limb brightening of the flow.}
   {}


   \maketitle

%
\section{Introduction}
The launching mechanism of relativistic extragalactic jets is a much debated and crucial topic for understanding these fascinating objects. There is currently a broad consensus on the fundamental role that the strong magnetic fields anchored in the accretion disk play in the process. However, to fully reproduce the observed properties of such magnetically-driven jets, e.g. Lorentz factors as high as $\sim$50 and opening angles as small as half a degree, is still a challenge for theorists. General relativistic MHD simulations \citep[e.g.][]{2011MNRAS.418L..79T} have shown that highly collimated, relativistic jets can be efficiently powered by the black hole's rotational energy extracted through the magnetic field, as described in the work of \cite{1977MNRAS.179..433B}. This scenario is supported by a recent observational study of a large sample of blazars \citep{2014Natur.515..376G}, indicating that the contribution of a spinning black hole is necessary to account for the estimated jet powers. 
Alternatively, the rotation of the accretion disk has been proposed as a viable power source for magnetic jet launching \citep{1982MNRAS.199..883B}. Predictions for the terminal Lorentz factor of such disk winds are diverse, with values ranging from mildly relativistic \citep[$\Gamma<3$,][] {2004ApJ...611..977M, 2005ApJ...620..878D} to relativistic \citep[$\Gamma\sim10$,][]{2007MNRAS.380...51K}.

A fundamental tool for discriminating between the two launching mechanisms is the direct imaging of the jet base through VLBI observations and the determination of its transverse size. This kind of study has been rarely performed because the jet base is usually not resolved. Moreover, owing to synchrotron opacity effects \citep{1981ApJ...243..700K}, the location of the VLBI core is often shifted at large distances, ${\sim}10^4-10^6$ Schwarzschild radii ($R_{\rm S}$), from the black hole \citep[e.g.][]{2008Natur.452..966M}. One exception is M\,87, for which a distance of only 14-23 $R_{\rm S}$ has been inferred for the 43 GHz core \citep{2011Natur.477..185H}. In the same source, VLBI pilot studies at 1.3 $\rm mm$ (with the Event Horizon Telescope) have determined a core size of ${\sim 5.5}$ $R_{\rm S}$ \citep{2012Sci...338..355D}, which represents an upper limit for the true size of the jet apex. This important measurement indicates that the jet of M\,87 is likely anchored close to the Innermost Stable Circular Orbit (ISCO). Performing such an analysis in other sources is clearly fundamental for testing the variety of possible scenarios. However, while future instruments with improved capabilities (e.g. phased ALMA) will certainly enrich the sample of suitable targets, this is currently quite limited. 

In this letter we present, for the first time, a detailed imaging of the jet base in \object{Cygnus\,A} through Global VLBI observations at 86 $\rm GHz$ (GMVA).
Among the powerful FR\,II radio galaxies, Cygnus\,A is one of the very few two-sided sources whose jet launching region can be probed with sufficient sensitivity at high radio frequencies.  At the source redshift (z=0.056)\footnote{$1$ $\rm{mas}{\sim} 1.084$ $\rm pc$, assuming a $\rm \Lambda CDM$ cosmology with ${\rm H_0} = 70.5$ $\rm {h}^{-1}$ $\rm km$ $\rm{s}^{-1}$ $\rm{Mpc}^{-1}$, $\rm \Omega_M=0.27$, $\rm {\Omega_{\Lambda}}=0.73$} and assuming a black hole mass of $2.5 \times 10^9 \rm M_{\odot}$  \citep{2003MNRAS.342..861T}, GMVA observations allow scales of the order of $10^2$ $R_{\rm S}$ to be probed. The large jet viewing angle enables a prominent counter-jet to be detected in this source, which represents a fundamental advantage when investigating the nuclear regions of a jet. In fact, since the transverse width can be determined continuously along the two-sided flow, the assumption-dependent back-extrapolations of the properties of the jet apex can be replaced by actual measurements. 

The following analysis complements a previous VLBI study at 43 $\rm GHz$ of the jet transverse profile and of the kinematic properties \citep[][hereafter B1]{2016A&A...585A..33B}. This revealed the presence of a pronounced transverse stratification, both in flux density (limb brightening) and bulk Lorentz factor ($1\lesssim\Gamma\lesssim2.5$), typical of a flow with spine-sheath structure. 
\begin{table}[width=\textwidth]
\caption{Log of observations at 86 GHz (3 mm) and characteristics of the clean maps shown in Fig. 1. Col. 1: Date of observation. Col. 2: Array. VLBA\_3mm indicates the part of the Very Long Baseline Array (all but the Hancock and St. Croix antennas) which is equipped with 86 GHz receivers; On - Onsala 20 m; Eb - Effelsberg 100-m; PdB - Plateau de Bure Interferometer; PV - IRAM 30-m atop Pico Veleta; Mh - Mets\"{a}hovi. Col. 3: Beam FWHM and position angle for untapered data with uniform weighting.}
\small
\centering
\begin{tabular}{c c c}
\hline
\hline
 Date & Array & Beam FWHM \\
 &&$\mathrm {[mas, deg]}$\\
\hline
08/05/2009 & VLBA\_{3mm}, On, &$0.12\times 0.05, -4.2^{\circ}$ \\
&Eb, PdB, PV &           \\
 10/10/2009 & VLBA\_3mm, On,&$0.10\times 0.04, -1.9^{\circ}$\\
&Eb, PdB, PV&\\
 07/05/2010 & VLBA\_3mm, On,&$0.11\times 0.04, -7.1^{\circ}$\\
&Eb, PdB, PV, Mh&\\
\hline
\end{tabular}\\
\end{table}

\section {Observations and VLBI images}
The GMVA data set comprises three epochs from observations in 2009/2010 with a cadence between five and seven months (Table 1). The duration of the observations was 15-16 hours, ${\sim}$8 of which were spent on Cygnus\,A. This, together with the large number of antennas employed and with the favourable declination of the source, enabled good uv-coverage. 
Data were recorded in single polarization mode with a recording rate of 512 {$\rm Mb/s$}, which corresponds to an observing bandwidth of 128 $\rm MHz$.
The correlation took place at the MPIfR in Bonn, Germany. The data reduction was carried out in AIPS \citep{1990apaa.conf..125G}. Aside from the standard calibration procedures, special care was devoted during the fringe fitting. First, as described in \cite{2012A&A...542A.107M}, the manual phase calibration was applied. Single sub-band phases and delays were calculated using one or more scans showing strong detections from the calibrator \object{J2015+3710}. Then, the global fringe fitting was performed \citep{1983AJ.....88..688S}. The amplitude calibration was based on the measured system temperatures and gain curves at each telescope, and it included a correction for atmospheric opacity. 

The imaging and self-calibration of amplitude and phase were performed in DIFMAP \citep{1994BAAS...26..987S}. 
The final clean maps obtained are presented in Fig. 1.  
The nuclear region is detected with a high signal-to-noise ratio S/N (between $\sim$1200-2000), with the brightest feature -- which we refer to as the ``core'' -- showing a prominent variability over the monitoring period. The total flux density varies between ${\sim}$0.65 and 1 $\rm Jy$ and is consistent with the measurement of ${\sim}1$ $\rm Jy$ provided by 86 $\rm GHz$ single-dish observations at IRAM \citep{2010ApJS..189....1A}.
\begin{figure}[!h]
\centering
 \includegraphics[trim=0cm 0cm 0cm 0cm, clip=true, width=0.5\textwidth]{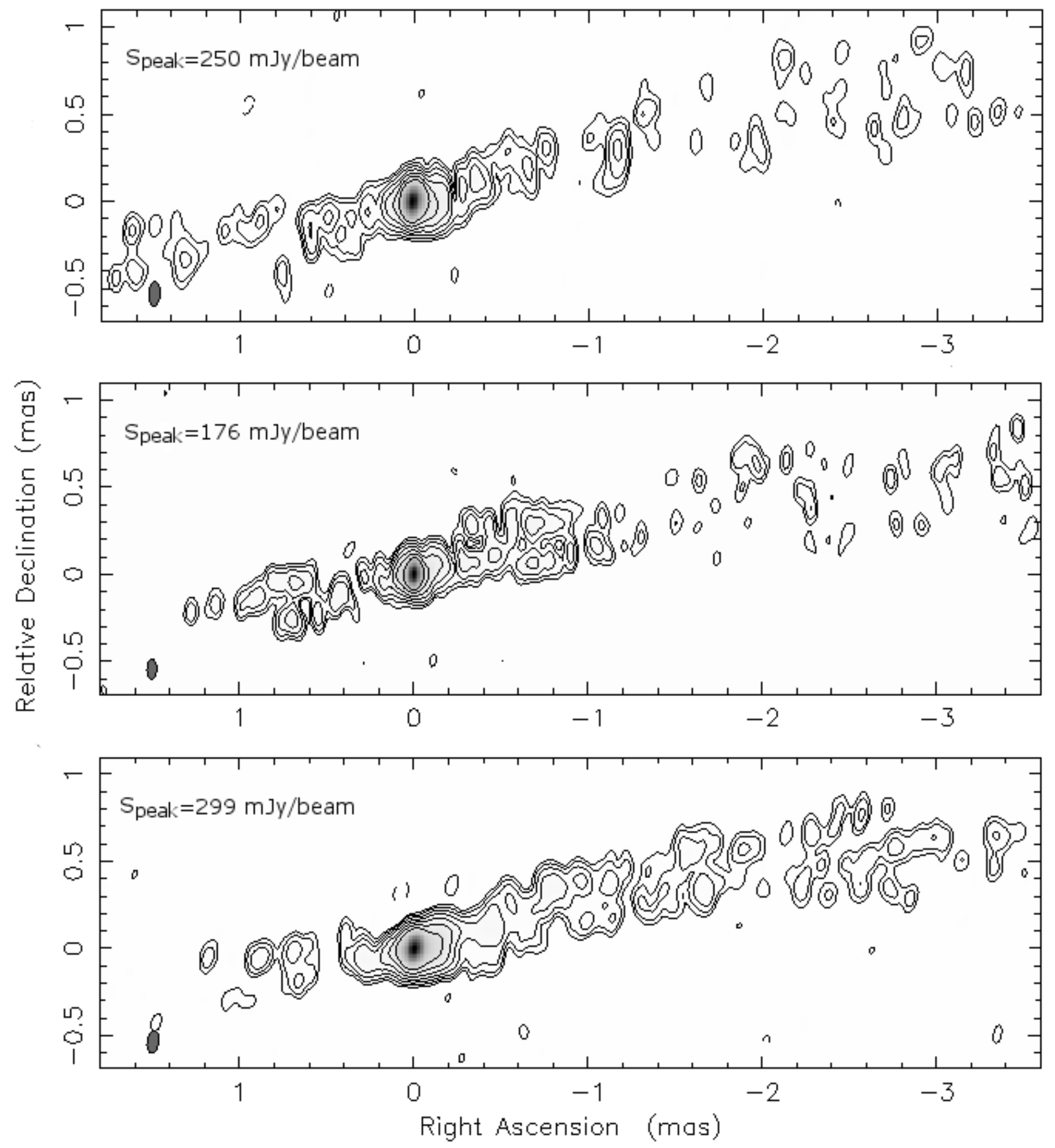}
 \caption{Clean maps of Cygnus\,A at 86 $\rm GHz$ from GMVA observations in 2009-2010. Contours represent isophotes at 0.6, 1.2, 2.4, 4.8, 9.6, 19.2, 38.4 $\rm mJy/beam$. The data weighting is uniform. The beam sizes are slightly larger compared to the values reported in Tab. 1 due to the application of a Gaussian taper of 0.1 at a uv-radius of 3000 $\rm M\lambda$. 
 The typical noise level in the images is ${\sim}0.15$ $\rm mJy/beam$. }
\end{figure}
The amplitude variation during the self-calibration procedure was not greater than $\sim$30\%, thus this percentage can be assumed as the uncertainty associated with the flux density measurements.

A visual inspection of the maps already shows the high morphological complexity of Cygnus\,A at this frequency and at this angular resolution. A prominent counter-jet is detected in all three epochs. The jet emission is relatively smooth -- i.e. there are no bright features besides the core -- but at the same time rich with sub-structures. It can also be noticed that, due to the low S/N in the faintest regions, the emission is not detected over the full jet cross-section in a single epoch. The jet structure can be recovered much better after stacking the three maps, as shown in Fig. 2. In this image, characterized by a reduced noise level of ${\sim}0.1$ $\rm mJy/beam$, the jet limb brightening already observed at 43 GHz also appears to be better defined, especially in the outer and faintest regions of the approaching jet. 
\begin{figure*}[!ht]
\centering
 \includegraphics[trim=0cm 0cm 0cm 0cm, clip=true, width=\textwidth]{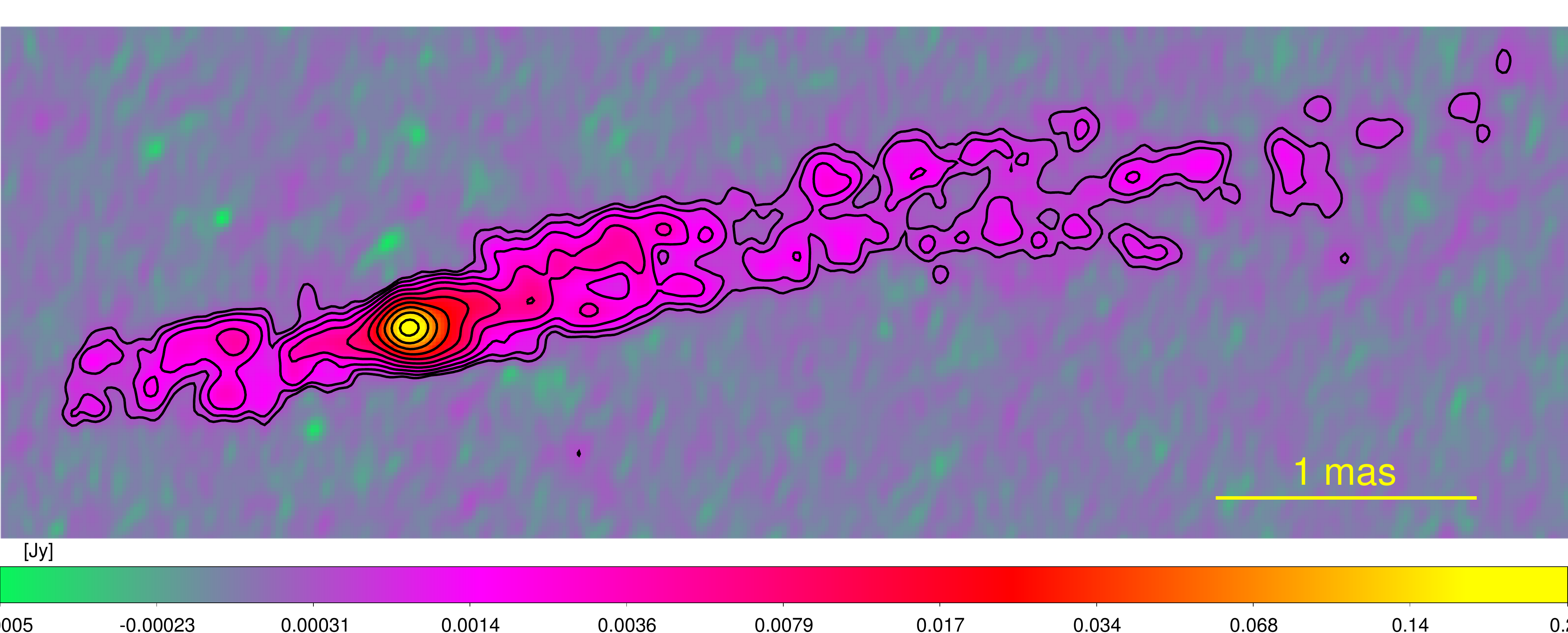}
 \caption{Stacked image of Cygnus A at 86 $\rm GHz$. Each single epoch was restored with a circular beam of 0.1 $\rm{mas}$. The alignment was based on the position of the peak of intensity. Contours represent isophotes at 0.4, 0.8, 1.6, 3.2, 6.4, 12.8, 51.6, 102.4, 204.8 $\rm mJy/beam$.}
\end{figure*}
\section{Analysis and discussion}
\subsection{Transverse width profile}
The analysis of the jet transverse structure in Cygnus\,A was first presented at the frequency of 43 $\rm GHz$ in B1. 
In this study it was shown that the flow emanates parabolically from the position of an emission gap located at ${\sim}0.2$ $\rm{mas}$ east of the brightest feature. 
A measurement of the jet transverse width was also provided for the nuclear region. However, the minimum size of ${\sim}30$ $\rm \mu as$ occurring in the vicinity of the faint emission gap was about seven times smaller than the beam size in the transverse direction. In this letter we exploit the improved resolution achieved by GMVA observations for obtaining a more solid estimate of the transverse width at the base of the flow.  
Given the better quality and the reduced noise level of the stacked image as compared to the single epochs, we focus our analysis on the former. 
The stacked image was restored with a circular beam of 0.1 $\rm mas$, which is approximately the natural resolution in the transverse direction (Table 1). Then, it was sliced pixel by pixel (1px=0.02 $\rm mas$) along the jet axis using the \textsc{AIPS} task \textsc{SLICE}. The transverse intensity distribution in each slice was finally fitted with a single Gaussian profile using the task \textsc{SLFIT}. 

Figure 3 shows the dependence of the de-convolved jet width on the distance from the core $z$ in the central ${\sim}1.5$ $\rm{mas}$ of the source. 
The distance $z$, also expressed in $R_{\bf S}$, was de-projected assuming a viewing angle $\theta=74.5^{\circ}$ (as calculated in B1). 
The error bars are conservatively assumed equal to one fifth of the measured FWHM. 
The minimum jet width $d_{\rm min}$ occurs for ${z=-0.02}$ $\rm{mas}$, where the measured transverse size is $d_{\rm min}=(51\pm22)$ $\rm \mu as$. This corresponds to about half of the nominal resolution in the transverse direction. However, as discussed by several authors \citep[e.g.][]{1989ASPC....6..213F, 2005astro.ph..3225L, 2005AJ....130.2473K}, the limiting resolution of a certain brightness distribution mainly depends on the S/N of the detection. Following \cite{2005astro.ph..3225L}, the resolution limit $d_{\rm lim}$ for the case of a Gaussian profile fitted to a brightness distribution with uniform weighting is
\begin{equation}
 d_{\rm lim}=\frac{4}{\pi}\,\sqrt{\pi\ln(2)b\ln\left(\frac{(S_{\rm p}/\sigma)}{(S_{\rm p}/\sigma)-1}\right)}
\end{equation}
where $S_{\rm p}$ is the peak flux density of the transverse intensity profile, $\sigma$ is the rms noise of the map and $b$  the equivalent circular beam (for uniform weighting)
expressed as a function of the major and minor beam axes, $b_{\rm maj}$ and $b_{\rm min}$, as $b=\sqrt{b_{\rm maj}\cdot b_{\rm min}}$.
The value of $d_{\rm lim}$ was calculated for each slice, and its profile as a function of $z$ is also reported in Fig. 3. Except for a faint area in the counter-jet (${z\sim-0.5}$ $\rm{mas}$), the measured width is everywhere well above the resolution limit, therefore the jet is resolved in the nuclear region.
\begin{figure}[!h]
\centering
  \includegraphics[trim=0cm 0cm 0cm 0cm, clip=true,  width=\linewidth]{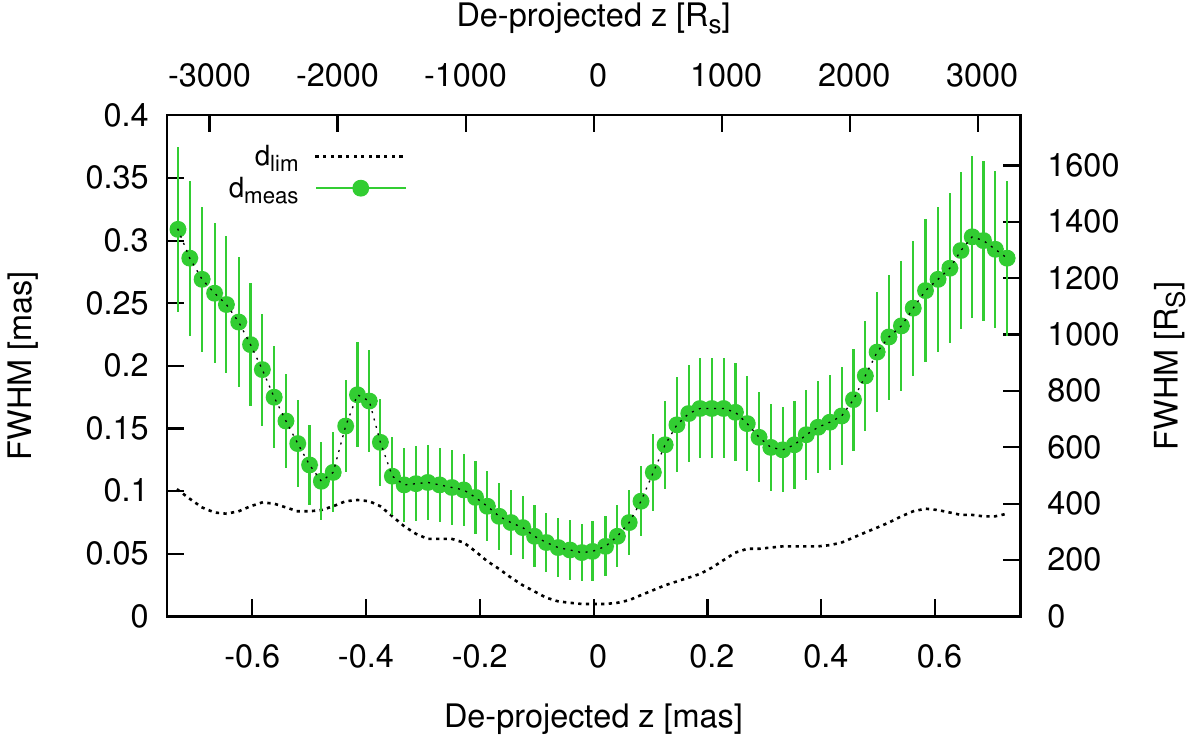}
 \caption{Transverse width profile at 86 $\rm GHz$ for the stacked image. Error bars for the width measurements (green points with label $d_{\rm meas}$) are assumed equal to 1/5 of each convolved FWHM. The dashed black line with label $d_{\rm lim}$ indicates the profile of the resolution limit.}
\end{figure}
\subsection{Minimum width and implications for the jet launching}
The minimum jet width $d_{\rm min}=(51\pm22)$ $\mu$as measured at 86 $\rm GHz$ translates into a linear size of $\sim(227\pm98)$ $R_{\bf S}$. 
Unlike the case of M\,87 \citep{2012Sci...338..355D}, the jet apex appears to be much wider than the ISCO radius, which varies between 1 and 9 $R_{\bf S}$ depending on the BH spin and on the kind of orbit (prograde or retrograde) of the BH/accretion disk system. These results suggest that at least part of the jet base in Cygnus\,A is anchored at large disk radii from the central BH, i.e. the outer regions of the accretion disk are involved in the jet launching process. 
Based on the sub-parsec scale kinematic study presented in B1, we infer that such a disk wind is characterized by mildly relativistic speeds, with the bulk Lorentz factor varying from $\Gamma\sim1$ at the jet edge to $\Gamma\sim2.5$ in sections of the flow closer to the jet axis. 

We point out, however, that our width measurement does not exclude the existence of both a narrower and an even broader component. 
Concerning the former, it should be recalled that in jets seen at large viewing angles Doppler de-boosting affects not only the receding side but also the approaching one. For sufficiently high speeds, the Doppler factor can be much smaller than one, so that the flow is completely invisible. In the case of Cygnus\,A, the gradual fading of the accelerating inner-jet components was directly observed in our kinematic analysis. This, together with the observed limb brightening, strongly indicate that an invisible faster flow exists in the central body of the jet. This spine may be driven either from the rotation of the inner regions of the accretion disk or by the spinning black hole itself. A hypothetical broader and diffuse disk wind, instead, is likely resolved out in these high resolution images. Indeed VLBI observations at 5 $\rm GHz$ \citep{1991AJ....102.1691C} support the existence of such a component. At this frequency, the jet appears as an almost cylindrical resolved flow with transverse width between 2 and 2.2 $\rm mas$, i.e. of the order of 9000 $R_{\bf S}$.  Although this large size may partially be the effect of scattering in the interstellar medium, there is an indication that the jet base extends to much larger disk radii than measured at 86 $\rm GHz$.
\subsection{Comparison with results at 43 GHz}
The discussion in the previous section has not quite addressed the problem of truly locating the central engine in Cygnus\,A. Given the roughly symmetric expansion of jet and counter-jet shown in Fig. 3, the most natural interpretation would be that the supermassive black hole is located in the vicinity of $z\sim0$. However, although the jet base is assumed to be the narrowest part of the jet in the simplest models, this may not necessarily be the case. An interesting argument comes from the comparison of the 86 $\rm GHz$ width profile with results at 43 $\rm GHz$. For the purposes of the comparison, the analysis was repeated at the two frequencies following the method described in Sect. 3.1 after convolving the maps with the same circular beam of 0.15 $\rm{mas}$. Results are presented in Fig. 4.
\begin{figure}[!h]
\centering
  \includegraphics[trim=0cm 0cm 0cm 0cm, clip=true, width=\linewidth]{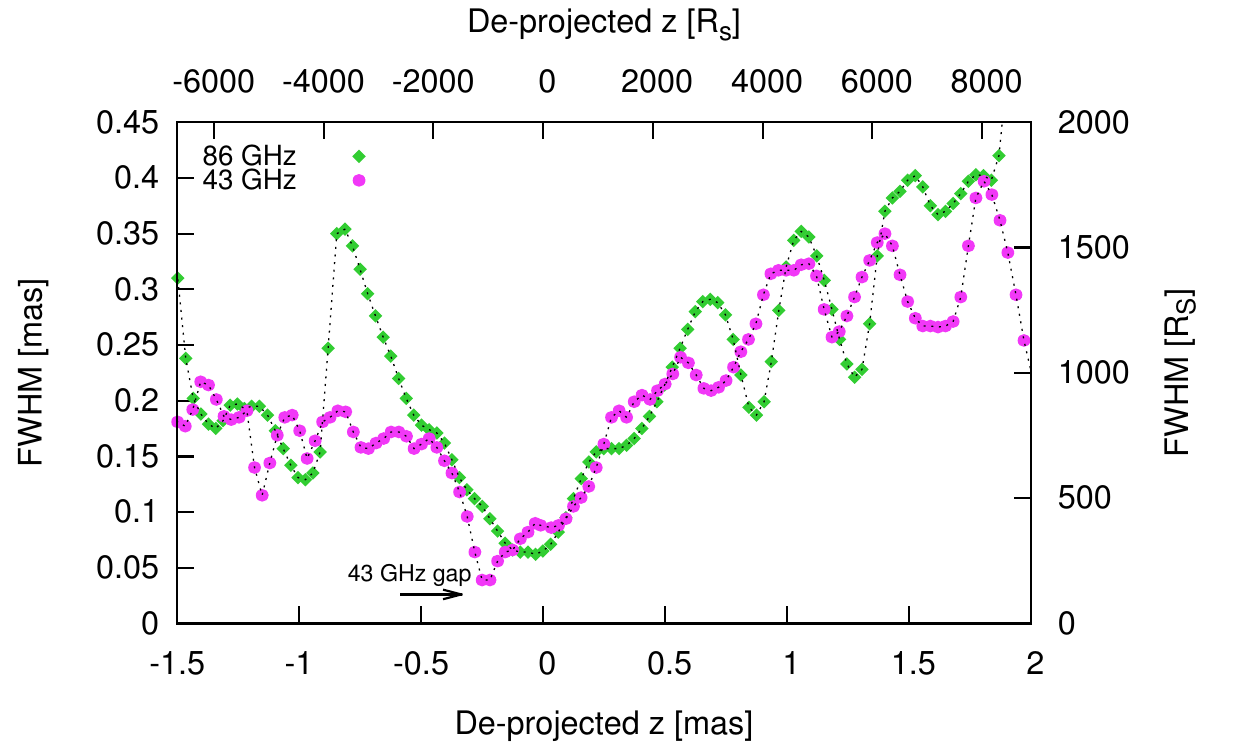}
 \caption{Width profiles at 43 and 86 $\rm GHz$. At 43 $\rm GHz$, the analysis was performed on the stacked image presented in B1, Fig. 2. The arrow indicates the position of the emission gap in the 43 $\rm GHz$ map. }
\end{figure}
We first notice that the dip in the 43 $\rm GHz$ profile at ${z\sim-0.25}$ $\rm{mas}$, i.e. in the vicinity of the gap of emission, is not observed at 86 $\rm GHz$.  
Correspondingly, the 86 $\rm GHz$ map (Fig. 3) does not show a pronounced emission gap.\footnote{In principle, this comparison may be affected by a misalignment of the maps due to an opacity shift of the cores. However, this is expected to be negligible at high frequencies, as confirmed by a cross-correlation of the jet parameters from which we infer a shift $<30$ $\rm\mu as$.}
The observed frequency dependence of the properties of this region supports the idea that the gap is the result of absorption, which becomes very reduced at 86 $\rm GHz$. If synchrotron opacity is responsible for the it, the gap most likely marks the true location of the central engine. A detailed spectral analysis using more sensitive data is in progress with the aim of clarifying the nature of this feature. Absorption may also explain another striking aspect of the width profiles in Fig. 4. While results at the two frequencies are quite consistent in the jet side, the 43 $\rm GHz$ counter-jet appears narrower than at 86 $\rm GHz$. The puzzling asymmetry between the opening angles of the approaching and receding side observed at 43 GHz in B1, almost disappears at 86 GHz. The existence of a parsec scale free-free absorber covering the counter-jet was suggested by several studies \citep{1998A&A...329..873K,2004evn..conf..155B, 2010A&A...513A..10S} and may well explain the observed features. The results presented in this letter hint at the presence of sub-pc scale material that is still highly opaque at 43 GHz but not at 86 $\rm GHz$.  
\section{Conclusions}
We have presented the first VLBI images of the radio galaxy Cygnus\,A at the frequency of 86 $\rm GHz$. The enhanced spatial resolution of these maps allowed us to investigate the transverse width profile at the onset of the two-sided flow. The jet base appears wide with a minimum width of ${\sim}(227\pm98)$ $R_{\bf S}$. This value is much higher than the ISCO radius, thus we infer that the emission is produced by a mildly relativistic, parabolically expanding disk wind. The existence of a central and faster jet, possibly driven by the black hole and invisible due to Doppler de-boosting, is however not excluded. It is, on the contrary, suggested by the kinematic properties and by the observed limb brightening of the flow. In the hypothesis that the \cite{1977MNRAS.179..433B} and \cite{1982MNRAS.199..883B} mechanisms act simultaneously in relativistic jets, giving rise to a spine-sheath structure of the flow \citep[e.g.][]{2007Ap&SS.311..281H, 2012RAA....12..817X}, we indeed expect the slower disk wind to always dominate the emission in radio galaxies and the BH-driven jet to be the main observable component in blazars due to differential boosting. The comparison of the width profiles at 43 and 86 $\rm GHz$ indicates that free-free opacity and/or synchrotron opacity are still strong at 43 $\rm GHz$ but not at 86 $\rm GHz$. Observing at high radio frequencies is therefore fundamental for unveiling the physical properties of jets at their onset. 
\begin{acknowledgements} 
We thank the referee for helpful comments. We also thank Andrei Lobanov and Vassilis Karamanavis for the useful discussions, and Michael Lindqvist for his support during the observations at Onsala.  This research has made use of data obtained with the Global Millimeter VLBI Array (GMVA), which consists of telescopes operated by the MPIfR, IRAM, Onsala, Metsahovi, Yebes and the VLBA. The data were correlated at the correlator of the MPIfR in Bonn, Germany. The VLBA is an instrument of the National Radio Astronomy Observatory, a facility of the National Science Foundation operated under cooperative agreement by Associated Universities, Inc.
 \end{acknowledgements}
\bibliographystyle{aa}
\bibliography{reference.bib}

\begin{thebibliography}{27}
\expandafter\ifx\csname natexlab\endcsname\relax\def\natexlab#1{#1}\fi

\bibitem[{{Agudo} {et~al.}(2010){Agudo}, {Thum}, {Wiesemeyer}, \&
  {Krichbaum}}]{2010ApJS..189....1A}
{Agudo}, I., {Thum}, C., {Wiesemeyer}, H., \& {Krichbaum}, T.~P. 2010, \apjs,
  189, 1

\bibitem[{{Bach} {et~al.}(2004){Bach}, {Krichbaum}, {Middelberg}, {Kadler},
  {Alef}, {Witzel}, \& {Zensus}}]{2004evn..conf..155B}
{Bach}, U., {Krichbaum}, T.~P., {Middelberg}, E., {et~al.} 2004, in European
  VLBI Network on New Developments in VLBI Science and Technology, ed.
  R.~{Bachiller}, F.~{Colomer}, J.-F. {Desmurs}, \& P.~{de Vicente}, 155--156

\bibitem[{{Blandford} \& {Payne}(1982)}]{1982MNRAS.199..883B}
{Blandford}, R.~D. \& {Payne}, D.~G. 1982, \mnras, 199, 883

\bibitem[{{Blandford} \& {Znajek}(1977)}]{1977MNRAS.179..433B}
{Blandford}, R.~D. \& {Znajek}, R.~L. 1977, \mnras, 179, 433

\bibitem[{{Boccardi} {et~al.}(2016){Boccardi}, {Krichbaum}, {Bach}, {Mertens},
  {Ros}, {Alef}, \& {Zensus}}]{2016A&A...585A..33B}
{Boccardi}, B., {Krichbaum}, T.~P., {Bach}, U., {et~al.} 2016, \aap, 585, A33

\bibitem[{{Carilli} {et~al.}(1991){Carilli}, {Bartel}, \&
  {Linfield}}]{1991AJ....102.1691C}
{Carilli}, C.~L., {Bartel}, N., \& {Linfield}, R.~P. 1991, \aj, 102, 1691

\bibitem[{{De Villiers} {et~al.}(2005){De Villiers}, {Hawley}, {Krolik}, \&
  {Hirose}}]{2005ApJ...620..878D}
{De Villiers}, J.-P., {Hawley}, J.~F., {Krolik}, J.~H., \& {Hirose}, S. 2005,
  \apj, 620, 878

\bibitem[{{Doeleman} {et~al.}(2012){Doeleman}, {Fish}, {Schenck}, {Beaudoin},
  {Blundell}, {Bower}, {Broderick}, {Chamberlin}, {Freund}, {Friberg},
  {Gurwell}, {Ho}, {Honma}, {Inoue}, {Krichbaum}, {Lamb}, {Loeb}, {Lonsdale},
  {Marrone}, {Moran}, {Oyama}, {Plambeck}, {Primiani}, {Rogers}, {Smythe},
  {SooHoo}, {Strittmatter}, {Tilanus}, {Titus}, {Weintroub}, {Wright}, {Young},
  \& {Ziurys}}]{2012Sci...338..355D}
{Doeleman}, S.~S., {Fish}, V.~L., {Schenck}, D.~E., {et~al.} 2012, Science,
  338, 355

\bibitem[{{Fomalont}(1989)}]{1989ASPC....6..213F}
{Fomalont}, E.~B. 1989, in Astronomical Society of the Pacific Conference
  Series, Vol.~6, Synthesis Imaging in Radio Astronomy, ed. R.~A. {Perley},
  F.~R. {Schwab}, \& A.~H. {Bridle}, 213

\bibitem[{{Ghisellini} {et~al.}(2014){Ghisellini}, {Tavecchio}, {Maraschi},
  {Celotti}, \& {Sbarrato}}]{2014Natur.515..376G}
{Ghisellini}, G., {Tavecchio}, F., {Maraschi}, L., {Celotti}, A., \&
  {Sbarrato}, T. 2014, \nat, 515, 376

\bibitem[{{Greisen}(1990)}]{1990apaa.conf..125G}
{Greisen}, E.~W. 1990, in Acquisition, Processing and Archiving of Astronomical
  Images, ed. G.~{Longo} \& G.~{Sedmak}, 125--142

\bibitem[{{Hada} {et~al.}(2011){Hada}, {Doi}, {Kino}, {Nagai}, {Hagiwara}, \&
  {Kawaguchi}}]{2011Natur.477..185H}
{Hada}, K., {Doi}, A., {Kino}, M., {et~al.} 2011, \nat, 477, 185

\bibitem[{{Hardee} {et~al.}(2007){Hardee}, {Mizuno}, \&
  {Nishikawa}}]{2007Ap&SS.311..281H}
{Hardee}, P., {Mizuno}, Y., \& {Nishikawa}, K.-I. 2007, \apss, 311, 281

\bibitem[{{Komissarov} {et~al.}(2007){Komissarov}, {Barkov}, {Vlahakis}, \&
  {K{\"o}nigl}}]{2007MNRAS.380...51K}
{Komissarov}, S.~S., {Barkov}, M.~V., {Vlahakis}, N., \& {K{\"o}nigl}, A. 2007,
  \mnras, 380, 51

\bibitem[{{K\"{o}nigl}(1981)}]{1981ApJ...243..700K}
{K\"{o}nigl}, A. 1981, \apj, 243, 700

\bibitem[{{Kovalev} {et~al.}(2005){Kovalev}, {Kellermann}, {Lister}, {Homan},
  {Vermeulen}, {Cohen}, {Ros}, {Kadler}, {Lobanov}, {Zensus}, {Kardashev},
  {Gurvits}, {Aller}, \& {Aller}}]{2005AJ....130.2473K}
{Kovalev}, Y.~Y., {Kellermann}, K.~I., {Lister}, M.~L., {et~al.} 2005, \aj,
  130, 2473

\bibitem[{{Krichbaum} {et~al.}(1998){Krichbaum}, {Alef}, {Witzel}, {Zensus},
  {Booth}, {Greve}, \& {Rogers}}]{1998A&A...329..873K}
{Krichbaum}, T.~P., {Alef}, W., {Witzel}, A., {et~al.} 1998, \aap, 329, 873

\bibitem[{{Lobanov}(2005)}]{2005astro.ph..3225L}
{Lobanov}, A.~P. 2005, ArXiv Astrophysics e-prints:0503225

\bibitem[{{Marscher} {et~al.}(2008){Marscher}, {Jorstad}, {D'Arcangelo},
  {Smith}, {Williams}, {Larionov}, {Oh}, {Olmstead}, {Aller}, {Aller},
  {McHardy}, {L{\"a}hteenm{\"a}ki}, {Tornikoski}, {Valtaoja}, {Hagen-Thorn},
  {Kopatskaya}, {Gear}, {Tosti}, {Kurtanidze}, {Nikolashvili}, {Sigua},
  {Miller}, \& {Ryle}}]{2008Natur.452..966M}
{Marscher}, A.~P., {Jorstad}, S.~G., {D'Arcangelo}, F.~D., {et~al.} 2008, \nat,
  452, 966

\bibitem[{{Mart{\'{\i}}-Vidal} {et~al.}(2012){Mart{\'{\i}}-Vidal}, {Krichbaum},
  {Marscher}, {Alef}, {Bertarini}, {Bach}, {Schinzel}, {Rottmann}, {Anderson},
  {Zensus}, {Bremer}, {Sanchez}, {Lindqvist}, \&
  {Mujunen}}]{2012A&A...542A.107M}
{Mart{\'{\i}}-Vidal}, I., {Krichbaum}, T.~P., {Marscher}, A., {et~al.} 2012,
  \aap, 542, A107

\bibitem[{{McKinney} \& {Gammie}(2004)}]{2004ApJ...611..977M}
{McKinney}, J.~C. \& {Gammie}, C.~F. 2004, \apj, 611, 977

\bibitem[{{Schwab} \& {Cotton}(1983)}]{1983AJ.....88..688S}
{Schwab}, F.~R. \& {Cotton}, W.~D. 1983, \aj, 88, 688

\bibitem[{{Shepherd} {et~al.}(1994){Shepherd}, {Pearson}, \&
  {Taylor}}]{1994BAAS...26..987S}
{Shepherd}, M.~C., {Pearson}, T.~J., \& {Taylor}, G.~B. 1994, in Bulletin of
  the American Astronomical Society, Vol.~26, Bulletin of the American
  Astronomical Society, 987--989

\bibitem[{{Struve} \& {Conway}(2010)}]{2010A&A...513A..10S}
{Struve}, C. \& {Conway}, J.~E. 2010, \aap, 513, A10

\bibitem[{{Tadhunter} {et~al.}(2003){Tadhunter}, {Marconi}, {Axon}, {Wills},
  {Robinson}, \& {Jackson}}]{2003MNRAS.342..861T}
{Tadhunter}, C., {Marconi}, A., {Axon}, D., {et~al.} 2003, \mnras, 342, 861

\bibitem[{{Tchekhovskoy} {et~al.}(2011){Tchekhovskoy}, {Narayan}, \&
  {McKinney}}]{2011MNRAS.418L..79T}
{Tchekhovskoy}, A., {Narayan}, R., \& {McKinney}, J.~C. 2011, \mnras, 418, L79

\bibitem[{{Xie} {et~al.}(2012){Xie}, {Lei}, {Zou}, {Wang}, {Wu}, \&
  {Wang}}]{2012RAA....12..817X}
{Xie}, W., {Lei}, W.-H., {Zou}, Y.-C., {et~al.} 2012, Research in Astronomy and
  Astrophysics, 12, 817

\end{thebibliography}
\end{document}